\documentclass[aps,pra,twocolumn,superscriptaddress,amsmath,amssymb,preprintnumbers,showpacs]{revtex4}

\usepackage{graphicx}
\usepackage{dcolumn}

\begin{document}


\title{Simulating quantum Brownian motion with single trapped ions}


\author{S. Maniscalco}
\affiliation{INFM, MIUR and Dipartimento di Scienze Fisiche ed
Astronomiche dell'Universit\`{a} di Palermo, via Archirafi 36,
90123 Palermo, Italy.} \email{sabrina@fisica.unipa.it}

\author{J.Piilo}
\affiliation{Department of Physics, University of Turku, FIN-20014
Turun yliopisto, Finland}
\affiliation{Helsinki Institute of
Physics, PL 64, FIN-00014 Helsingin yliopisto, Finland}

\author{F. Intravaia}
\affiliation{Laboratoire Kastler Brossel,\footnote{\'{E}cole
Normale Sup\'{e}rieure, Centre National de la Recherche
Scientifique, Universit\'{e} Pierre et Marie Curie.} Case 74, 4
place Jussieu, F-75252 Paris Cedex 05, France.}

\author{F. Petruccione}
\affiliation{Physikalisches Institut,
Albert-Ludwigs-Universit\"{a}t, Hermann-Herder Stra\ss e 3,
D-79104 Freiburg im Breisgau, Germany}

\affiliation{School of Pure and Applied Physics, University of
KwaZulu-Natal, Durban 4041, South Africa}

\author{A. Messina}
\affiliation{INFM, MIUR and Dipartimento di Scienze Fisiche ed
Astronomiche dell'Universit\`{a} di Palermo, via Archirafi 36,
90123 Palermo, Italy.}

\date{\today}

\begin{abstract}
We study the open system dynamics of a harmonic oscillator coupled
with an artificially engineered reservoir. We single out the
reservoir and system variables governing the passage between
Lindblad type and non-Lindblad type dynamics of the reduced
system's oscillator. We demonstrate the existence of conditions
under which virtual exchanges of energy between system and
reservoir take place. We propose to use a single trapped ion
coupled to engineered reservoirs in order to simulate quantum
Brownian motion.
\end{abstract}


\maketitle

\section{Introduction}
The dynamics of closed systems may be calculated exactly by
solving directly the Schr\"{o}dinger equation. In realistic
physical conditions, however, the system one is interested in is
coupled to  its surrounding. In this case the dynamics of the
total closed system can be extremely complicated. For this reason,
since the very early days of quantum mechanics, a huge deal of
attention has been devoted to the study of the dynamics of open
quantum systems \cite{petruccionebook}.

Nowadays the interest in such a broad field has notably increased
mainly for two reasons. On the one hand experimental advances in
the coherent control of single or few atom systems have paved the
way to the realization of the first basic elements of quantum
computers, c-not \cite{cnot} and phase quantum gates
\cite{qgates}. Moreover, the first quantum cryptographic
\cite{crypt} and quantum teleportation \cite{telep} schemes have
been experimentally implemented. These technological applications
rely on the persistence of quantum coherence. Thus, understanding
decoherence and dissipation arising from the unavoidable
interaction between the system and its surrounding is necessary in
order to implement real-size quantum computers
\cite{NISTqcomputer} and quantum technologies. On the other hand,
one of the most debated aspects of quantum mechanics, namely the
quantum measurement problem, can be interpreted in terms of
environment induced decoherence \cite{Zurek}. According to this
interpretation the emergence of the classical world from the
quantum world can be seen as a decoherence process due to the
interaction between system and environment \cite{giulini}.

A paradigmatic model of the theory of open systems is a harmonic
oscillator linearly coupled with a reservoir modelled as an
infinite set of non interacting oscillators. Indeed this model is
central in many physical contexts, e.g., quantum field theory
\cite{cohen}, quantum optics
\cite{petruccionebook,carmichael,buzeck}, and solid state physics
\cite{weiss}.

In order to describe quantitatively and qualitatively how the
reservoir affects the system dynamics one needs to make some
assumptions on its nature and properties. Some of these
properties, as for example the temperature of the reservoir, can
be experimentally measured. Other parameters, as the reservoir
spectral density or the system-reservoir coupling, are assumed on
the basis of physical reasonableness and deduced by the comparison
with experimental data. In this sense the model is a
phenomenological one.

The importance of the damped harmonic oscillator is also due to
the fact that it is one of the few exactly solvable non-trivial
systems. In fact the Heisenberg equations of motion for the total
system (oscillator plus reservoir) can easily be solved. The
solution of the Heisenberg equations of motion is indeed the most
straightforward method for the description of the dynamics of the
expectation values of observables of interest, e.g. the mean
energy of the system. An example showing the easiness and
conceptual transparency of the method is given in
\cite{petruccionebook}.

Another way to describe the time evolution of the system is to
look at the reduced density matrix which is obtained by tracing
the density matrix of the total system over the environmental
degrees of freedom. This procedure is motivated by the fact that,
in general, one is interested only in the dynamics of the system
and not in the state of the reservoir. An exact master equation
for the reduced density matrix can be formulated and exactly
solved
\cite{Haake,Feynman,Caldeira,Hu,Ford,Grabert,PRAsolanalitica}.

During the last decade huge advances in laser cooling and trapping
experimental techniques have made it possible to confine
harmonically a single ion and cool it down to very low
temperatures where purely quantum manifestations begin to play an
important role \cite{trapreview}. A single laser cooled ion is
theoretically equivalent to a particle moving in a harmonic
potential, whose center of mass motion is quantized as a harmonic
oscillator. Such a system is a unique experimental system since it
approximates very well a closed system \cite{trapreview}. Indeed
unwanted dissipation, which in this case manifests itself as a
heating process depopulating the vibrational ground state of the
ion, is negligible for times much longer than the usual times in
which experiments take place \cite{heatingNIST,simuNIST}.
Moreover, arbitrary states of the ion motion can be prepared and
coherently manipulated using proper laser pulses
\cite{statesNIST,statesBlatt}. Even extremely fragile states as
the Schr\"{o}dinger cat states have been realized and detected
\cite{schcat}. Quite recently, by using multiple ions in a linear
Paul trap, experiments on quantum non-locality have been performed
\cite{qnonlocality} and many-particle entangled states have been
realized \cite{4entanglement}. Furthermore cold trapped ions are
the favorite candidates for a physical implementation of quantum
computers \cite{cnot,NISTqcomputer}.

The aim of this paper is to study the interaction of a quantum
harmonic oscillator with engineered reservoirs. In the context of
trapped ions it is possible not only to engineer experimentally an
\lq\lq artificial\rq\rq~reservoir but also to synthesize both its
spectral density  and the coupling with the system oscillator
\cite{engineerTzero,engineerNIST}. This makes it possible to think
of new types of experiments aimed at testing the predictions of
fundamental models as the one of quantum Brownian motion (QBM) (or
its high $T$ limit: the famous Caldeira-Leggett model
\cite{Caldeira}).

The mean vibrational quantum number of the ion, also called
heating function \cite{heatingNIST}, is the central quantity we
investigate. We study the heating dynamics of the single
oscillator in correspondence to different reservoir characteristic
parameters. We analyze the influence of the variations of the
engineered reservoir parameters, e.g., the cut-off frequency of
the reservoir spectral density, on the heating process. The idea
of looking at the variation of the open system dynamics induced by
the changes of the relevant reservoir parameters is in fact rather
unusual. Up to recently, indeed, only dissipation and/or
decoherence due to interaction with the \lq\lq natural\rq\rq~
reservoir were studied
\cite{petruccionebook,carmichael,weiss,gardiner,Haake,lindenberg}.

We compare our analytic results for the observable quantities of
interest, as the heating function, with the non-Markovian wave
function (NMWF) simulations  \cite{petruccionebook,Breuer99a},
finding a very good agreement. The main result of the paper is the
experimental proposal for observing new features of quantum
Brownian motion with single trapped ions. We demonstrate that with
currently available technology a new regime of the open system
dynamics, characterized by virtual phonon exchanges between the
system and the reservoir, may be explored.

The paper is organized as follows. In Sec.~\ref{sec:exact}  we
introduce the Master Equation for QBM and its solution obtained by
means of a superoperatorial approach. In Sec.~\ref{sec:results} we
study the behavior of the heating function for different values of
the reservoir parameters. In Sec.~\ref{exptechnique} we review the
basic ingredients of the experimental procedures for engineering
artificial amplitude reservoir, as the one we study in the paper,
and for measuring the heating function. In Sec.~\ref{expQBM} we
describe our experimental proposal for revealing
non-Markovian dynamics of a quantum Brownian particle, simulated
with a single trapped ion. Finally in Sec.~\ref{sec:conclusion}
conclusions are presented.

\section{Exact dynamics of a quantum Brownian particle}
\label{sec:exact}
\subsection{Generalized Master Equation}\label{sec:rwaME}
The dynamics of a harmonic oscillator linearly coupled with a
quantized reservoir, modelled as an infinite chain of quantum
harmonic oscillators, may be described exactly by means of a
generalized Master Equation of the form
\cite{petruccionebook,Hu,EPJRWA,annals}
\begin{eqnarray}
\frac{d \rho_S(t)}{dt} = \frac{1}{i \hbar} \mathbf{H}_{0}^S
\rho_S(t)
 -  \Big[ \Delta(t) (\mathbf{X}^S)^2 - \Pi(t) \mathbf{X}^S
\mathbf{P}^S \nonumber \\
 - \frac{i}{2} r(t) (\mathbf{X^2})^S + i \gamma(t) \mathbf{X}^S
\mathbf{P}^{\Sigma} \Big]\rho_S(t). \label{QBMme}
\end{eqnarray}
We indicate with $\mathbf{X}^{S(\Sigma)}$ and
$\mathbf{P}^{S(\Sigma)}$ the commutator (anticommutator) position
and momentum superoperators respectively and with
$\mathbf{H}_{0}^S$ the commutator superoperator relative to the
system Hamiltonian. The time dependent coefficients appearing in
the Master Equation can be written, to the second order in the
coupling strength, as follows
\begin{eqnarray}
\Delta(t)&=& \int_0^t \kappa(\tau) \cos (\omega_0 \tau) d \tau,
\label{delta} \\
\gamma(t) &=& \int_0^t \mu(\tau) \sin (\omega_0
\tau) d \tau, \label{gamma} \\
\Pi(t) &=& \int_0^t \kappa(\tau) \sin (\omega_0
\tau) d \tau, \label{pi} \\
r(t)&=& 2 \int_0^t \mu(\tau) \cos (\omega_0 \tau) d \tau
\label{erre},
\end{eqnarray}
where
\begin{equation}
\kappa(\tau)= \alpha^2 \langle \{ E(\tau),E(0)\} \rangle,
\label{kappa}
\end{equation}
and
\begin{equation}
\mu(\tau)= i \alpha^2 \langle [ E(\tau),E(0)] \rangle, \label{mu}
\end{equation}
are the noise and dissipation kernels, respectively. In the
previous equations we indicate with $\alpha$ the system--reservoir
coupling constant, with $\omega_0$ the frequency of the system
oscillator and with $E$ the generalized reservoir position
operator.

The Master Equation (\ref{QBMme}) is local in time, even if
non-Markovian. This feature is typical of all the generalized
Master Equations derived by using the time-convolutionless
projection operator technique \cite{petruccionebook,Breuer99a} or
equivalent approaches such as the superoperatorial one presented
in \cite{PRAsolanalitica,royer}.

The time dependent coefficients appearing in Eq.~(\ref{QBMme})
contain all the information about the short time system-reservoir
correlation. The coefficient $r(t)$ gives rise to a time dependent
renormalization of the frequency of the oscillator. The term
proportional to $\gamma(t)$ is a classical damping term while the
coefficients $\Delta(t)$ and $\Pi(t)$ are diffusive terms.

In what follows we  study the time evolution of the heating
function $\langle n (t) \rangle$ with $n$ quantum number operator.
The dynamics of $\langle n (t) \rangle$ depends only on the
diffusion coefficient $\Delta(t)$ and on the classical damping
coefficient $\gamma(t)$ \cite{PRAsolanalitica}. Furthermore, the
quantum number operator $n$ belongs to a class of observables not
influenced by a secular approximation which takes the Master
Equation (\ref{QBMme}) into the form \cite{PRAsolanalitica,grab}
\begin{eqnarray}
\frac{ d \rho_S(t)}{d t}\!\!&=&\!\! \frac{\Delta(t) \!+\! \gamma
(t)}{2} \left[2 a \rho_S(t) a^{\dag}- a^{\dag} a \rho_S(t)  -
\rho_S(t) a^{\dag} a \right]
\nonumber \\
& &\!\!\frac{\Delta(t) \!-\! \gamma (t)}{2} \left[2 a^{\dag}
\rho_S(t) a - a a^{\dag} \rho_S(t) - \rho_S(t) a a^{\dag}
 \right]. \nonumber \\
 \label{MERWA}
\end{eqnarray}
For this reason, in order to calculate the exact time evolution of
the heating function, one can use the solution of the approximated
Master Equation (\ref{MERWA}). In the previous equation we have
introduced the bosonic annihilation and creation operators
$a=\left( X+i P \right)/\sqrt{2}$ and $a^{\dag}=\left( X-i P
\right)/\sqrt{2}$, with $X$ and $P$ dimensionless position and
momentum operator. Note that the above Master Equation is of
Lindblad type as far as the coefficients $ \Delta(t) \pm \gamma
(t)$ are positive \cite{Maniscalco04a}.

\subsection{Analytic solution}

The superoperatorial Master Equation (\ref{QBMme}) can be exactly
solved by using specific algebraic properties of the
superoperators \cite{PRAsolanalitica}. The solution for the
density matrix of the system is derived in terms of the Quantum
Characteristic Function (QCF) $\chi_t(\xi)$ at time $t$, defined
through the equation \cite{cohen}
\begin{equation}
\label{sdef} \rho_S(t)=\frac{1}{2\pi}\int \chi_t(\xi)\:
e^{\left(\xi a^{\dag}-\xi^* a\right)} d^2\xi.
\end{equation}
It is worth noting that one of the advantages of the
superoperatorial approach is the relative easiness in calculating
the analytic expression for the mean values of observables of
interest by using the relation
\begin{eqnarray}
\langle a^{\dag m} a^n \rangle = \left. \left(\frac{d}{d
\xi}\right)^m \left(- \frac{d}{d \xi^*}\right)^n
e^{|\xi|^2/2}\chi(\xi) \right|_{\xi=0}. \label{a}
\end{eqnarray}
The exact analytic expression for the time evolution of the
heating function can be obtained from the approximated RWA
solution. In the RWA the QCF is \cite{PRAsolanalitica}
\begin{equation}
\chi_t (\xi)=e^{- \Delta_{\Gamma}(t) |\xi|^2} \chi_0 \left[ e^{-
\Gamma (t)/2} e^{-i \omega_0 t} \xi  \right], \label{chit}
\end{equation}
with $\chi_0$ QCF of the initial state of the system. The
quantities $\Delta_{\Gamma}(t)$ and $\Gamma(t)$ appearing in
Eq.~(\ref{chit})
 are defined in terms of the diffusion and dissipation
 coefficients $\Delta(t)$ and $\gamma(t)$ respectively  as follows
\begin{eqnarray}
\Gamma(t)&=& 2\int_0^t \gamma(t_1)\:dt_1, \label{Gamma} \\
\Delta_{\Gamma}(t) &=& e^{-\Gamma(t)}\int_0^t
e^{\Gamma(t_1)}\Delta(t_1)dt_1 \label{DeltaGamma}.
\end{eqnarray}
Eq.~(\ref{chit}) shows that the QCF is the product of an
exponential factor, depending on both the diffusion $\Delta(t)$
and the dissipation $\gamma(t)$ coefficients, and a transformed
initial QCF. The exponential term accounts for energy dissipation
and is independent of the initial state of the system. Information
on the initial state is given by the second term of the product,
the transformed initial QCF.

Having in mind Eq.~(\ref{chit}) and using Eq.~(\ref{a}), one gets
the following expression for the heating function
\begin{equation}
\label{hf} \langle n(t) \rangle = e^{-\Gamma(t)} \langle n(0)
\rangle + \frac{1}{2} \left(  e^{-\Gamma(t)}  - 1\right) +
\Delta_{\Gamma}(t).
\end{equation}
The asymptotic long time behavior of the heating function is
readily obtained by using the Markovian stationary values for
$\Delta(t)$ and $\gamma(t)$. For a  thermal reservoir, one gets
\begin{equation}
\langle n (t) \rangle =e^{- \Gamma t} \langle n(0) \rangle +
n(\omega_0) \left( 1- e^{- \Gamma t}\right). \label{nM}
\end{equation}
In the next section we will discuss in detail the dynamics of the
heating process and we will show the changes in the short time
dynamics due to the variations of typical reservoir parameters
such as its temperature and cut-off frequency.

\section{Non-Markovian dynamics of Lindblad and non--Lindblad type}\label{sec:results}
In a previous paper we have presented a theory of heating for a
single trapped ion interacting with a natural reservoir able to
describe both its short time non-Markovian behavior and the
asymptotic thermalization process \cite{letteranostra}. In this
paper we focus instead on the case of interaction with engineered
reservoirs. In the trapped ion context, it is experimentally
possible to engineer \lq \lq artificial\rq\rq~  reservoirs and
couple them  to the system in a controlled way. Since the coupling
with the natural reservoir is negligible for long time intervals
\cite{trapreview,heatingNIST}, this allows to test fundamental
models of open system dynamics as the one for QBM we are
interested in. In a sense this extends the idea of using trapped
ions for simulating the closed dynamics of quantum optical systems
\cite{simuNIST,feynman2,simuwinel} to the possibility of
simulating the dynamics of an ubiquitous open system as the damped
harmonic oscillator. In particular, by using the analytic
solution, one can look for ranges of the relevant parameters of
both the reservoir and the system in correspondence of which
deviations from Markovian dissipation become experimentally
observable.


In the experiments on artificially engineered amplitude reservoirs
\cite{engineerNIST} the high temperature condition $\hbar \omega_0
/ KT \ll 1$ is always satisfied. For this reason here we
concentrate on this regime of the parameters. For an Ohmic
reservoir spectral density with Lorentz-Drude cut-off
\begin{equation}
J(\omega)= \frac{2  \omega}{\pi} \
\frac{\omega_c^2}{\omega_c^2+\omega^2}, \label{spectrald}
\end{equation}
the dissipation and damping coefficients $\gamma(t)$ and
$\Delta(t)$, appearing in the Master Equation (\ref{MERWA}), to
second order in the coupling constant, take the form
\begin{equation}
\gamma (t)\! \!=\!\! \frac{\alpha^2 \omega_0 r^2}{1+r^2} \Big[1
\!-\! e^{- \omega_c t} \cos(\omega_0 t) \! - r e^{- \omega_c t}
\sin( \omega_0 t )  \Big], \label{gammasecord}
\end{equation}
and
\begin{eqnarray}
\Delta(t) &=& 2 \alpha^2 k T \frac{r^2}{1+r^2} \left\{ 1
- e^{-\omega_c t} \left[ \cos (\omega_0 t)\right.\right. \nonumber \\
&-&  (1/r) \left. \left. \sin (\omega_0 t )\right] \right\},
\label{deltaHT}
\end{eqnarray}
with $r=\omega_c/\omega_0$. Equation (\ref{deltaHT}) has been
derived in the high $T$ limit, while $\gamma(t)$ does not depend
on temperature. Comparing Eq.~(\ref{deltaHT}) with
Eq.~(\ref{gammasecord}), one notices immediately that in the high
temperature regime, $\Delta(t) \gg \gamma(t)$. Having this in mind
it is easy to prove that the heating function, given by
Eq.~(\ref{hf}), may be written in the following approximated form
\begin{equation}
\label{hfhighT} \langle n(t) \rangle \simeq
\Delta_{\Gamma}(t),
\end{equation}
where we have assumed that the initial state of the ion is its
vibrational ground state, as it is actually the case at the end of
the resolved sideband cooling process
\cite{statesNIST,statesBlatt,cooling}. For times much bigger than
the reservoir correlation time $\tau_R= 1/ \omega_c$ the
asymptotic behavior of Eq.~(\ref{hfhighT}) is given by
Eq.~(\ref{nM}). This equation gives evidence for a second
characteristic time of the dynamics, namely the thermalization
time $\tau_T= 1/ \Gamma $, with $\Gamma= \alpha^2 \omega_0 r^2
/(r^2+1)$. The thermalization time depends both on the coupling
strength and on the ratio $r=\omega_c / \omega_0$ between the
reservoir cut-off frequency and the system oscillator frequency.
Usually, when one studies QBM, the condition  $r \gg 1$, which
corresponds to a natural flat reservoir, is assumed. In this case
the thermalization time is simply inversely proportional to the
coupling strength. For an \lq\lq out of resonance\rq\rq~engineered
reservoir with $r \ll 1$, $\tau_T$ is notably increased and
therefore the thermalization process is slowed down.

A further approximation to the heating function of
Eq.~(\ref{hfhighT}) can be obtained for times $t \ll \tau_T $:
\begin{eqnarray}
\label{hfhighTappr} \langle n(t) \rangle \!\! &\simeq& \!\!
\int_0^t \Delta(t_1) dt_1  = \frac{2 \alpha^2 KT}{\omega_c}
\frac{r^2}{(r^2+1)^2}
\Big\{ \omega_c t (r^2+1) \\
&-&  \!\! (\! r^2 \! - \! 1 \!) \! \left[ 1-e^{- \omega_c t} \!
\cos (\omega_0 t)\right] \!-\! r e^{- \omega_c t} \! \sin
(\omega_0 t) \! \Big\}.
\end{eqnarray}
This approximation shows a clear connection between the sign of
the diffusion coefficient $\Delta(t)$ and the time evolution of
the heating function before thermalization. The diffusion
coefficient is indeed the time derivative of the heating function.
We remind that, since $\Delta(t)\gg \gamma (t)$ for the case
considered here, whenever $\Delta(t) > 0$ the Master Equation
(\ref{MERWA}) is of Lindblad type, whilst the case $\Delta(t) < 0$
corresponds to a non-Lindblad type Master Equation. From
Eq.~(\ref{hfhighTappr}) one sees immediately that while for $\Delta(t)
> 0$ the heating function grows monotonically, when
$\Delta(t)$ assumes negative values it can decrease and present
oscillations.

To better understand such a behavior we study in more details the
dynamics for three exemplary values of the ratio $r$ between the
reservoir cut-off frequency and the system oscillator frequency:
$r \gg 1$, $r=1$ and $r\ll 1$. As we have already noticed the
first case corresponds to the assumption commonly done when
dealing with natural reservoir while the last case corresponds to
an engineered \lq\lq out of resonance\rq\rq~  reservoir.

For $r\gg 1$ the diffusion coefficient,  given by
Eq.~(\ref{deltaHT}), is positive for all $t$ and $r$ since
\begin{equation}
\Delta (t) \varpropto 1- e^{- \omega_c t} \cos(\omega_0 t) \ge 0.
\end{equation}
Therefore the Master Equation is always of Lindblad type and the
heating function grows monotonically from its initial null value.
Equation (\ref{hfhighTappr}) shows that, for times $t \ll \tau_R $
and for $r \gg 1$, $\langle n(t) \rangle \simeq (\alpha^2 \omega_c
kT ) t^2$, that is the initial non-Markovian behavior of the
heating function is quadratic in time.

For $r=1$, a similar behavior  is observed since also in this case
$\Delta(t)$ is positive at all times.

Finally, in the case $r \ll 1$, Eq.~(\ref{deltaHT}) shows that, if
$r$ is sufficiently small, $\Delta (t)$ oscillates acquiring also
negative values. It is worth noting, however, that the long time
asymptotic value of $\Delta(t)$ is always positive. Whenever the
diffusion coefficient is negative, the heating function decreases,
so the overall heating process is characterized by oscillations of
the heating function. The decrease in the population of the ground
state of the system oscillator, after an initial increase due to
the interaction with the high $T$ reservoir, is due to the
emission and subsequent reabsorption of the same quantum of
energy. Such an event is possible since the reservoir correlation
time $\tau_R = 1/ \omega_c$ is now much longer than the period of
oscillation $\tau_s = 1/ \omega_0$. We underline that, although
the Master Equation in this case is not of Lindblad type, it
conserves the positivity of the reduced density matrix. This of
course does not contradict the Lindblad theorem since the
semigroup property is clearly violated for the reduced system
dynamics \cite{petruccionebook}.

\section{Experimental techniques}\label{exptechnique}
This section
gives
a brief review of the experimental procedures for engineering
artificial reservoirs and for measuring the heating function of
the trapped ion. Starting from
the careful analysis of the recent experiments presented in this
section,
we will describe, in the Sec.~\ref{expQBM},
our experimental
proposal for simulating QBM with single trapped ions.

\subsection{Engineering Reservoirs}\label{engres}

Let us begin discussing the technique used  to engineer an
artificial reservoir coupled to a single trapped ion. Reference
\cite{engineerNIST} presents recent experimental results showing
how to couple a properly engineered reservoir with a quantum
oscillator, namely  the quantized center of mass (c.m.) motion of
the ion. These experiments aim at measuring the decoherence of a
quantum superposition of coherent states and Fock states due to
the presence of the reservoir. Several types of engineered
reservoirs are demonstrated, e.g., thermal amplitude reservoirs,
phase reservoirs, and zero temperature reservoirs.

A high $T$ amplitude reservoir is obtained by applying a random
electric field $\vec{E}$ whose spectrum is centered on the axial
frequency $\omega_z / 2 \pi =11.3$MHz  of oscillation of the ion
\cite{engineerNIST}. The trapped ion motion couples to this field
due to the net charge $q$ of the ion: $H_{int}= -q \vec{x} \cdot
\vec{E}$, with $\vec{x}=(X,Y,Z)$ displacement of the  c.m. of the
ion from its equilibrium position. Remembering that $\vec{E}
\propto \sum_i \vec{\epsilon}_i (b_i + b^{\dag}_i)$, with $b_i$
and $b^{\dag}_i$ annihilation and creation operators of the
fluctuating field modes, and that $X \propto \left( a +a^{\dag}
\right)$ one realizes that this coupling is equivalent to the
bilinear one assumed to derive Eq.~(\ref{QBMme}).

The random electric field is applied to the endcap electrodes
through a network of properly arranged low pass filters limiting
the \lq\lq natural\rq\rq~environmental noise but allowing
deliberately large applied fields to be effective. This type of
drive simulates an infinite-bandwidth amplitude reservoir
\cite{engineerNIST}. It is worth stressing that, for the times of
duration of the experiment, namely $\Delta t=20 \mu$s, the heating
due to the natural reservoir is definitively negligible
\cite{engineerNIST}.

The reservoir considered in our paper is a thermal reservoir with
spectral distribution given by
\begin{eqnarray}
I(\omega) &=& J(\omega) [n(\omega)+1/2] \nonumber \\
&=& \frac{ \omega}{\pi} \frac{\omega_c^2}{\omega_c^2+\omega^2}
\coth{(\omega/K T)} , \label{eq:I}
\end{eqnarray}
where Eq.~(\ref{spectrald}) has been used. For high $T$,
Eq.~(\ref{eq:I}) becomes
\begin{equation}
I(\omega) = \frac{2 K T}{\pi}
\frac{\omega_c^2}{\omega_c^2+\omega^2}.
\end{equation}
The infinite-bandwidth amplitude reservoir realized in the
experiments corresponds to the case $\omega_c \rightarrow \infty$
in the previous equation. Therefore, for high $T$, the reservoir
discussed in the paper can be realized experimentally by filtering
the random field, used in the experiments for simulating an
infinite-bandwidth reservoir, with a Lorentzian shaped low pass
filter at frequency $\omega_c$. The change of the ratio $r$ thus
would be accomplished simply by changing the low pass filter.

\subsection{Measurements of the heating
function}\label{measurement}

In this subsection we focus on two experimental methods for
measuring the heating function of a single trapped ion. The first
method is based on the asymmetry in the sideband motional spectrum
of the ion and it has been used in \cite{heatingNIST} for
measuring the process of thermalization of an ion, initially
cooled down to its ground vibrational state, due to the
interaction with the natural reservoir. The same method is used in
\cite{statesBlatt} for measuring the cooling dynamics of an ion
subjected to sideband cooling lasers. The second technique allows
to measure the populations of the vibrational density matrix of
the ion, from which the heating function can be obtained. This
last method has been used in \cite{engineerNIST} in the case of
interaction with an artificial amplitude reservoir.

For both techniques the first step is the preparation of the
initial vibrational and electronic ground state, $\vert n=0, -
\rangle \equiv \vert n=0 \rangle \otimes \vert - \rangle$,
obtained by laser cooling and optical pumping to the state $\vert
- \rangle$. The mean vibrational number is then measured after
fixed delay time intervals. During the delay
artificial
noise,
which simulates the amplitude reservoir, may be applied. In this
way, the time evolution of $\langle n (t) \rangle $ is obtained.

Let us begin with the first technique. At each fixed delay time,
the ion is in its electronic ground state $\vert - \rangle$ and in
a certain vibrational state. A laser pulse, tuned to a vibrational
sideband, is then used to transfer the population to the upper
electronic level $\vert + \rangle$. After this, by means of an
electron shelving technique, the electronic state of the ion is
detected in order to check wether a transition to $\vert +
\rangle$ has occurred or not. Repeating this procedure one gets
the electronic excited state occupation probability $P_+$. The
amplitude of the blue and red vibrational sidebands is defined as
the probability of making a transition $\vert - \rangle
\rightarrow \vert + \rangle$ due to  a laser pulse tuned to the
blue or red sideband, respectively, and therefore is given by
$P_+$. This quantity depends on the mean occupation number
$\langle n \rangle$. For $\vert n = 0 \rangle$, only the blue
sideband can be exited while the red one is absent. In general the
asymmetry in the amplitude of the $k$th red $(I^k_{red})$ and blue
$(I^k_{blue})$ vibrational sidebands, allows to extract $\langle n
\rangle$ \cite{heatingNIST}:
\begin{equation}
I^k_{red} = \left(\frac{\langle n \rangle}{\langle n \rangle +
1}\right)^k I^k_{blue}.
\end{equation}

A  limitation of this method is given by off-resonant excitation
via the carrier transition. If the driving field is tuned to the
first lower vibrational sideband, in the resolved sideband
condition $\omega_0 \gg \Omega$, with $\Omega$ Rabi frequency of
the laser pulse, processes involving off-resonant transitions go
as $(\Omega/ \omega_0)^2$. In order to have a sensible measurement
of the heating function, the population due to off-resonant
transitions have to be much smaller than the scale over which
$\langle n (t) \rangle$ varies.

We now focus on the second experimental method for measuring the
heating function. This method actually allows to measure the
diagonal elements $P_n$ of the vibrational density matrix and it
has been used to observe their decay due to the interaction with
an artificial amplitude reservoir, as the one described in
previous subsection. For this type of reservoir and in  the
experimental conditions of \cite{engineerNIST}, the time evolution
of $P_n (t)\equiv \rho_{nn}(t)$ is well approximated by the law
\begin{eqnarray}
&& \rho_{nn}(t) \simeq \frac{1}{1+ \bar{n}\gamma t} \sum_{j=0}^n
\left( \frac{\bar{n}\gamma t}{1+ \bar{n} \gamma t} \right)^j
\left( \frac{1}{1+ \bar{n} \gamma t} \right)^{2n-2j} \nonumber \\
&& \sum_{l=0}^{\infty}\!\! \left( \frac{\bar{n}\gamma t}{1+
\bar{n} \gamma t}\right)^l \!\!\!\left(\begin{array}{c}
      n+l-j \\
      n-j \end{array} \right) \!\!\left(\begin{array}{c}
      n \\
      j \end{array} \right)\!\! \rho_{n+l-j,n+l-j}(0), \nonumber \\
 \label{PnNist}
\end{eqnarray}
where the phenomenological parameters $\bar{n}$ and $\gamma$ are
the mean reservoir quantum number and the heating rate,
respectively. Equation (\ref{PnNist}) is valid under the
assumptions of high $T$ reservoir and for times much smaller than
the thermalization time, $\gamma t \ll 1$. Such conditions turn
out to be verified in the experiments described in
\cite{engineerNIST}. In this experimental situation, the Markovian
behaviour of the heating function, before thermalization takes
place, is simply given by $\langle n (t) \rangle = \bar{n} \gamma
t$. Note that in the limit $r\gg 1$, corresponding to an infinite
bandwidth Markovian reservoir, Eq.~(\ref{hfhighTappr}) becomes
$\langle n(t) \rangle \simeq 2 \alpha^2 KT t / \pi$. Comparing
this expression of the heating function with the one of the
experiments one can identify $\bar{n} = KT/ \omega_0$ and $\gamma
= 2 \alpha^2 \omega_0 /\pi$. It is worth underlining, however,
that according to Eq.~(\ref{PnNist}), only the product $\bar{n}
\gamma$ may be deduced from the experimental data.

In order to describe our experimental proposal for simulating QBM,
it is important to look in more detail at the  procedure used in
\cite{engineerNIST} to obtain the time evolution of the
populations $P_n$. This
allows to deduce from the experimental data the characteristic
parameters of the amplitude reservoir used in the experiments. In
\cite{engineerNIST}, after preparing a superposition of Fock
states, the amplitude noise is applied for a fixed time of
$\bar{t} =3 \mu$s after which the populations $P_n$ are measured.
In order to measure $P_n$ the ion is irradiated with a pair of
Raman beams tuned to the first blue sideband for various probe
times $t_p$, and $P_- (t_p)$ is measured from the fluorescence
signal. From the $P_-(t_p)$ data the populations $P_n (\bar{t})$
are finally extracted with a single value decomposition
\cite{engineerNIST}. In order to observe the time evolution of
$P_n$ one can proceed in two equivalent ways. Either one changes
the interval of time $\bar{t}$ during which the amplitude noise is
applied, or one fixes $\bar{t}$ and varies the variance of the
applied noise $\langle V^2 \rangle$. In fact, as shown in
\cite{engineerNIST}, the variance of the random noise used to
simulate an amplitude reservoir is $2 \langle V^2 \rangle =
\bar{n} \gamma t$. Practically, increasing the fluctuations of the
random electric field applied at the trap electrodes is equivalent
to an increase in the heating function $\langle n \rangle =
\bar{n} \gamma t$. Since $P_n(t)$, as given by
Eq.~(\ref{hfhighTappr}), depends only on $\bar{n} \gamma t$, one
can obtain the time evolution of the populations simply by
changing $\langle V^2 \rangle$. This is the method used in
\cite{engineerNIST} for measuring the populations, so, in fact, in
the experiment $P_n(\langle V^2 \rangle)$ is recorded. In
principle, by using the relation $\bar{n} \gamma = \langle V^2
\rangle / \bar{t}$, with $\bar{t}= 3 \mu$s, one could directly
obtain the characteristic parameter of the reservoir from the
value of the  noise voltage applied to the trap electrodes.
However, unknown geometrical factors in converting the voltage
to variations in the secular frequency prevent a direct
comparison. If we indicate with $\langle V^2 \rangle_{appl} = c
\langle V^2 \rangle$ the fluctuations of the voltage applied to
the electrodes, with $c \in \mathbb{R}$,  fitting the experimental
data according to the theoretical law given by Eq.~(\ref{PnNist}),
allows to extrapolate the factor $c$ and therefore $\bar{n}
\gamma$. It is not difficult to show that, in the experiment on
the decay (heating) of a Fock state due to interaction with
engineered amplitude reservoir (see Fig.~[15] of
\cite{engineerNIST}), $c\simeq10$ and hence, for $\bar{t}=3 \mu$s
and $\langle V^2 \rangle = 0.25 V^2$, $\bar{n}\gamma \simeq 0.84
\cdot 10^{7}$Hz. Note that, in the experiment, $\langle V^2
\rangle$ is varied from $0$ to $0.3 V^2$.

At this point we are ready to describe our experimental proposal
for observing the non-Markovian dynamics of the heating function
and,
in general, for simulating the dynamics of a quantum
Brownian particle.

\section{Experimental proposal for simulating QBM}\label{expQBM}
It is well known that non-Markovian features usually occur in the
dynamics for times $t \ll \tau_R = 1/ \omega_c$. In general, since
$\omega_c \gg \omega_0$ and typically $\omega_0 \simeq 10^{7}$Hz
for trapped ions, this means that deviations from the Markovian
dynamics appear for times $t \ll 0.1 \mu$s. This is the reason why
the initial quadratic behaviour of the heating function is not
observed in the experiments, wherein the typical time scales go
from $1$ to $100 \mu$s.

A way to force non-Markovian features to appear is to \lq
detune\rq~the trap frequency  from the reservoir spectral density.
This corresponds, for example, to the case in which $r=\omega_c /
\omega_0 =0.1$. In this case the reservoir correlation time is
bigger than the period of oscillation of the ion and this leads to
the oscillatory behaviour of the heating function predicted by
Eq.~(\ref{hfhighTappr}) for $r \ll 1$. Under this condition
$\tau_c =1 \mu$s, and therefore the non-Markovian features show up
in the time evolution and can be measured. Detuning the trap
frequency from the reservoir, however, decreases the effective
coupling between the system and the environment and, for this
reason, in order to obtain values of the heating function big
enough to be measured we need to increase either the coupling
constant $\alpha^2$, which correspond to an increase in the
intensity of the voltage applied to the electrodes, or the
strength of the fluctuations $\langle V^2 \rangle$, which
correspond to an increase in the effective temperature of the
reservoir.

\begin{figure}
\includegraphics[width=8 cm,height=7 cm]{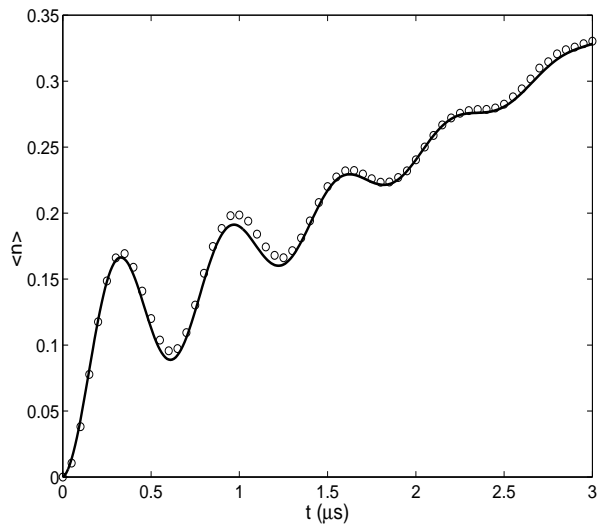}
\caption{\label{qbmexperiment} Time evolution of the heating
function for $2 \alpha^2 KT / \pi =  0.84 \cdot 10^{9}$Hz,
$\omega_c=1$MHz and $r=0.1$.
Solid line is the analytical and circles the simulation result.}
\end{figure}

Let us look in more detail to Eq.~(\ref{hfhighTappr}). For $r\ll
1$ this equation becomes
\begin{eqnarray}
\langle n(t) \rangle &\simeq& \frac{2 \alpha^2 KT}{\pi \omega_c}
r^2 \left\{ \omega_c t + \left[ 1- e^{-\omega_c t}\cos(\omega_0 t)
\right] \right. \nonumber \\
&& \left. - r e^{-\omega_c t} \sin (\omega_0 t) \right\}.
\label{nexp}
\end{eqnarray}
In the comparison with the experiment done in previous section we
have seen that the front factor $2 \alpha^2 KT / \pi = \bar{n}
\gamma \simeq 0.84 \cdot 10^{7}$Hz. Increasing of two order of
magnitude this front factor, and for $r=0.1$, Eq.~(\ref{nexp})
predicts the behaviour for the heating function shown in Fig.
\ref{qbmexperiment}.
We believe that for this range of $\langle n \rangle$ and of times
the oscillations of the heating function could be experimentally
measurable. For example, with the first technique described in
subsection \ref{measurement}, using $\Omega=10^{6}$Hz and for
$\omega_0=10^7$Hz, the ground state population transferred to the
excited level due to off resonant excitation is of the order of
$10^{-2}$, which is one order of magnitude smaller than the
variation of the heating function we want to measure (see Fig.
\ref{qbmexperiment}). Also the other  method described in previous
subsection seems to be enough accurate to reveal the oscillatory
behaviour of the heating function in the conditions here examined.
However, it is worth noting that, since the heating function does
not depend now on the dimensionless variable $\bar{n} \gamma t$,
but rather on $\omega_c t$, the time evolution can be obtained
only by varying the duration of the time of application of the
amplitude reservoir. This is not equivalent to a change in the
applied voltage fluctuations, as it was in the Markovian case
discussed in subsection \ref{measurement}.

Summarizing, in order to observe the virtual exchanges of phonons
between the system and the reservoir, leading to the oscillations
of the heating function, one needs to increase of two order of
magnitude the coefficient $2 \alpha^2 KT / \pi = \bar{n} \gamma$.
This can be done either increasing the intensity or increasing the
fluctuations of the applied noise, or combining an increase in the
intensity with an increase in the fluctuations. Moreover one needs
to use a low pass filter for the applied noise, as described in
subsection \ref{engres}, having cut off frequency $\omega_c=0.1
\omega_0$.

\begin{figure}
\includegraphics[width=8 cm,height=7 cm]{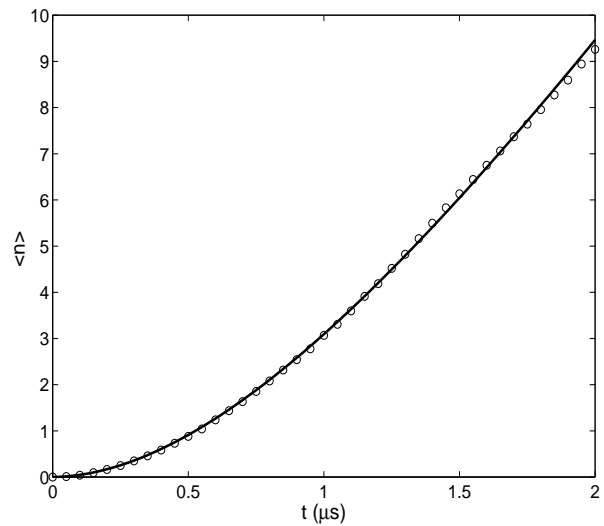}
\caption{\label{qbmexpquad} Time evolution of the heating function
for $2 \alpha^2 KT / \pi =  0.84 \cdot 10^{9}$Hz,
$\omega_c=1$MHz, and $r=10$. Solid line is the analytical and
circles the simulation result.}
\end{figure}

We now examine briefly the conditions for which the quadratic
behaviour of the heating function, could be observed. We remind
that this is the case in which $r \gg 1$ and the time evolution of
the  density matrix is of Lindblad-type. In view of the
considerations done at the beginning of this section, in order to
reveal non-Markovian dynamics in a time scale of $1-100 \mu$s we
need to have $\tau_R \ge 1 \mu$s, that is, for $r=10$, $\omega_0
\le 0.1$MHz. This means that one actually needs a \lq
loose\rq~trap. For example, for $\omega_0 = 100$kHz, with the same
applied noise used in \cite{engineerNIST}), i.e.~$\bar{n} \gamma =
0.84 \cdot 10^{7}$Hz, the time evolution of the heating function
is the one shown in Fig~\ref{qbmexpquad}.

If one wants to perform a fast measurement of $\langle n \rangle$,
e.g.~using the first method and assuming $\Omega = 10^6$Hz, the
small value of the trap frequency makes it difficult not only to
implement one of the two methods for measuring the heating
function, but also to reach the initial ground state since the
sidebands are not clearly resolved, and therefore resolved
sideband cooling technique cannot be applied. It is worth
stressing, however, that contrarily to the case of a natural
reservoir, which is always \lq in action\rq, in the case of an
engineered reservoir one can switch off the applied noise after a
certain delay time and, assuming that the effect of the natural
reservoir is negligible, measure the heating function without the
severe requirement of big values of $\Omega$. If one assumes that
after an amplitude noise pulse the state of the ion does not
change, then it is not necessary to perform a fast measurement of
$\langle n \rangle$. Therefore one can work with smaller values of
$\Omega$, such that $\Omega / \omega_0 \ll 1$.

Concluding, while measuring the quadratic behaviour of the heating
function ($r\gg 1$) could be a more challenging task from the
experimental point of view, revealing the oscillatory
non-Markovian behaviour ($r\ll 1$) appears to us in the grasp of
the experimentalists, in the conditions we have analyzed in this
section.

\section{Conclusions}\label{sec:conclusion}
In this paper we have studied
the dynamics of
a single harmonic oscillator coupled to a quantum reservoir at
generic temperature $T$. In our analysis we have used both the analytic
solution for the
 reduced density matrix and the NMWF method.

We have paid special attention to the non-Markovian heating
dynamics  typical of short times. In this regime the system time
evolution is influenced by correlations between the system and the
reservoir. For certain values of the system and reservoir
parameters, virtual exchanges of energy between the system and its
environment become dominant. These virtual processes strongly
affect the short time dynamics and are responsible for the
appearance of oscillations in the heating function (non-Lindblad
type dynamics).

Extending the ideas of using trapped ions for simulating quantum
optical systems, we have proposed to simulate QBM with single
trapped ions coupled to artificial    reservoirs. We have
carefully analyzed the possibility of revealing, by using present
technologies, the non-Markovian dynamics of a single trapped ion
interacting with an engineered reservoir and we have underlined
the conditions under which non-Markovian features become
observable.

\section{Acknowledgements}
The authors gratefully acknowledge Heinz-Peter Breuer for helpful
comments and stimulating discussions. J.P. acknowledges financial
support from the Academy of Finland (project no.50314) and the
Finnish IT center for Science (CSC) for computer resources. J.P.
and F.P. thank the University of Palermo for the hospitality. S.M.
acknowledges Finanziamento Progetto Giovani Ricercatori anno 1999
Comitato 02 for financial support and the Helsinki Institute of
Physics for the hospitality.

\end{document}